\documentclass[11pt]{article}
\usepackage{epsfig}
\usepackage{graphicx}
\usepackage{amsmath}
\usepackage{amssymb}

\begin{document}

\title{Establishing the astrophysical origin \\
of a signal in a neutrino telescope}

\author{Paolo Lipari}

\maketitle
\begin{abstract}
Recently the IceCube collaboration has reported the observation
of 28 contained events with a visible energy in the 
interval between $6 \times 10^{13}$~eV and $1.5 \times 10^{15}$~eV, and
has argued that this detection is 
evidence, with a statistical significance of more than four standard deviations,
for the existence of an astrophysical neutrino flux that accounts 
for a large fraction of the events.
In this work we analyze the arguments
that allow to identify a component of 
astrophysical origin in the high energy neutrino flux separating it from
atmospheric neutrinos.
An astrophysical origin for a large fraction of the IceCube 
contained events is the simplest and most natural explanation 
of the data but, conservatively,
an atmospheric origin cannot yet be entirely ruled out. 
This ambiguity should  soon be resolved.
\end{abstract}

\section{Introduction}
\label{sec:introduction}
Recently the IceCube detector has reported 
\cite{icecube-ICRC} the observation
of 28 ``contained events'' where an energy larger
than approximately 60~TeV (a minimum number of 6000 photo electrons)
is visible in the detector.
In the analysis the most external PMT's of the telescope are used as a veto
to select events where the energy deposition
is  confined to the inner part 
of the detector volume.
The selection cuts determine a fiducial mass of approximately 400~Mton.
The observation of the two highest energy events 
(with a deposited energy of $1.04 \pm 0.16$ and
 $1.14 \pm 0.17$~PeV) had been reported previously 
\cite{Aartsen:2013bka}
from  a search for contained events performed with a higher threshold.

The IceCube collaboration has estimated  
that the  expected number of events
(for the data taking period of 662 days)
is $N_{\rm expected} = 10.6^{+5.0}_{-3.6}$ 
in the absence  of an astrophysical component.
This estimate has a contribution  of $6.0 \pm 3.4$ events  
associated to down--going cosmic ray showers 
(this muon background was calculated extrapolating from data 
taken with a smaller fiducial volume and more stringent veto),
and a contribution of $4.6 \pm 1.2$ events from standard atmospheric neutrinos
(that is neutrinos created in the decay of charged pions and kaons).
The contribution of neutrinos from charm decay is estimated
as 1.5~events, and this is included in the error for $N_{\rm expected}$
and accounts  for its asymmetry ($+5.0$ versus $-3.6$ events).

According to the IceCube analysis,
the excess in the  observed number of  contained events 
can be explained assuming
the existence of an isotropic flux  of neutrinos that is approximately 
equal for the six neutrino types.
This flux 
(for each flavor and summing over neutrinos and anti--neutrinos)
has  the form:
\begin{equation}
E_\nu^2 \; \phi_{\nu_\alpha} (E) \simeq (1.2 \pm 0.4) 
\times 10^{-8} ~{\rm GeV}/({\rm cm}^2{\rm s}\; {\rm sr})~.
\label{eq:flux-evidence}
\end{equation}
This flux  cannot extend to energies larger
than approximately 2~PeV, because of the non--detection
of events with energy larger than $\simeq 1$~PeV.
The flux in equation (\ref{eq:flux-evidence}) is not a unique solution,
and softer spectra are also consistent with the data.

Note that  most of the contained events are due to the charged 
current interactions of electron and tau (anti)--neutrinos.
In the case of electron (anti)--neutrinos the visible energy
is equal  to  the incident neutrino energy.  For tau (anti)--neutrinos
a fraction of the  energy escapes the detector  
since the $\tau^\mp$ created
in the interactions always decay  into a neutrino.
The  observed rate of  charged current interactions of muon (anti)--neutrino
is  suppressed because of  the containement requirement.
A smaller number of events is  generated
by  neutral current  interactions, where however  a  large fraction
of the  energy escapes with the final state neutrinos.

The  detection of high energy astrophysical neutrinos has been a 
goal pursued for decades with  telescopes of always increasing
sensitivity, and is unquestionably a  result of 
profound significance.
Not surprisingly the IceCube results
have immediately attracted considerable interest, and several authors
have already commented on the possible 
implications of such an astrophysical neutrino flux.

In this work I take a very ``conservative'' attitude, and 
reanalyse the hypothesis  that the data 
can be explained  only with  atmospheric neutrinos.
This is motivated by the consideration that results of great importance
must be analysed with great critical attention.

The problem of the identification of an
astrophysical component in the observed fluxes of neutrinos
has  been already discussed extensively in the past.
The obvious and unquestionable method to determine the astrophysical 
origin of a set of neutrino events is the observation of 
a structure in their angular distribution in celestial coordinates.
Any such anisotropy would be immediately 
unambiguous proof that at least a fraction of 
the observed neutrinos originates from astrophysical 
sources outside the solar system. 
It has however been predicted 
(see for example \cite{Gaisser:1994yf,Lipari:2008zf})
that the first evidence for high energy astrophysical neutrinos could
emerge in the form of an isotropic flux. This is the simple consequence
of the fact that neutrinos can traverse the entire universe 
with negligible absorption, and the sum of all
unresolved extragalactic sources
(most of them faint and very distant) should form an isotropic flux
that is likely to be largest contribution to the neutrino sky.

The signal observed by IceCube is consistent with such an
isotropic angular distribution 
(even if the data shows some intriguing hints of 
anisotropy that have unavoidably generated speculations). 

Atmospheric neutrinos produced in pion and kaon decay
have a strong and characteristic zenith angle
dependence due to the fact that the decay probability
of the parent mesons is enhanced in inclined showers
that develop higher in the atmosphere where the air density is lower.
On the other hand prompt (that is generated in charm decay) neutrinos,
in the energy range considered, are also  to a very good approximation,
isotropic.

In the absence of the  unambigouos signature  of a
structure in the sky, the  astrophysical origin
of a component of  the neutrino flux can however be 
obtained from a study of the flavor composition and energy spectrum 
of the  neutrinos.

This work is organized as follows,
in the the next section we make a few comments on the angular distribution
of the neutrino fluxes and of the IceCube contained events.
In section~\ref{sec:muons}  we discuss the importance and 
sensitivity  of neutrino induced muons.
In section~\ref{sec:flavor}  we discuss the 
flavor compoition of the fluxes.
Sections~\ref{sec:nucleon-flux} and~\ref{sec:charm} 
discuss the  main uncertainties in  the calculation   of the atmospheric
neutrino  fluxes, namely the estimate of the primary cosmic ray (CR) 
nucleon flux, and the properties of the charm cross section, and 
discuss under which assumptions it possible to attribute the
rate of contained  events observed bu IceCube  to prompt atmospheric  
neutrinos.
The last section gives a summary and an outlook.

\section{Angular distribution}
\label{sec:angular-distribution}
Of the 28 events reported in \cite{icecube-ICRC}, 24 are down--going
and only 4 are upgoing (with a ratio $N_{\rm down}/N_{\rm up} \simeq 6$).

\begin{figure} [ht]
\begin{center}
\includegraphics[width=14.5cm]{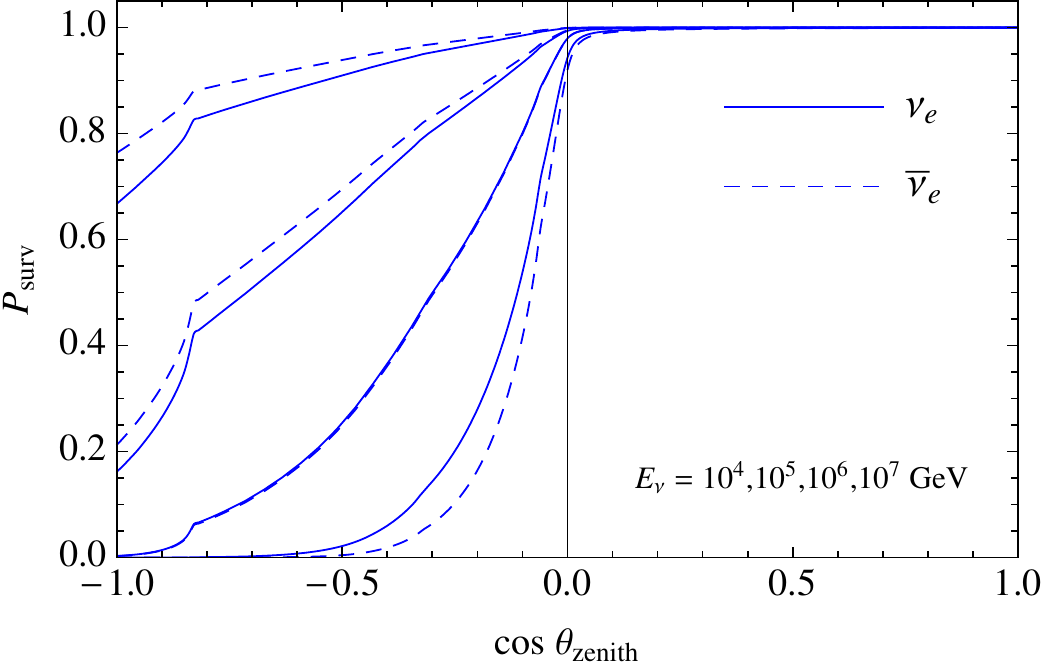}
\end{center}
\caption {\footnotesize
Absorption probability for $\nu_e$ (solid lines) and
$\overline{\nu}_e$  (dashed lines) of different energies
($E_\nu = 10^{4}$, 
$10^{5}$, 
$10^{6}$ and 
$10^{7}$~GeV) plotted as a function
of zenith angle.
The calculation was performed using the
neutrino cross section of 
Gandhi et al. \cite{Gandhi:1998ri}.
\label{fig:absorption}
 }
\end{figure}

\begin{figure} [ht]
\begin{center}
\includegraphics[width=14.5cm]{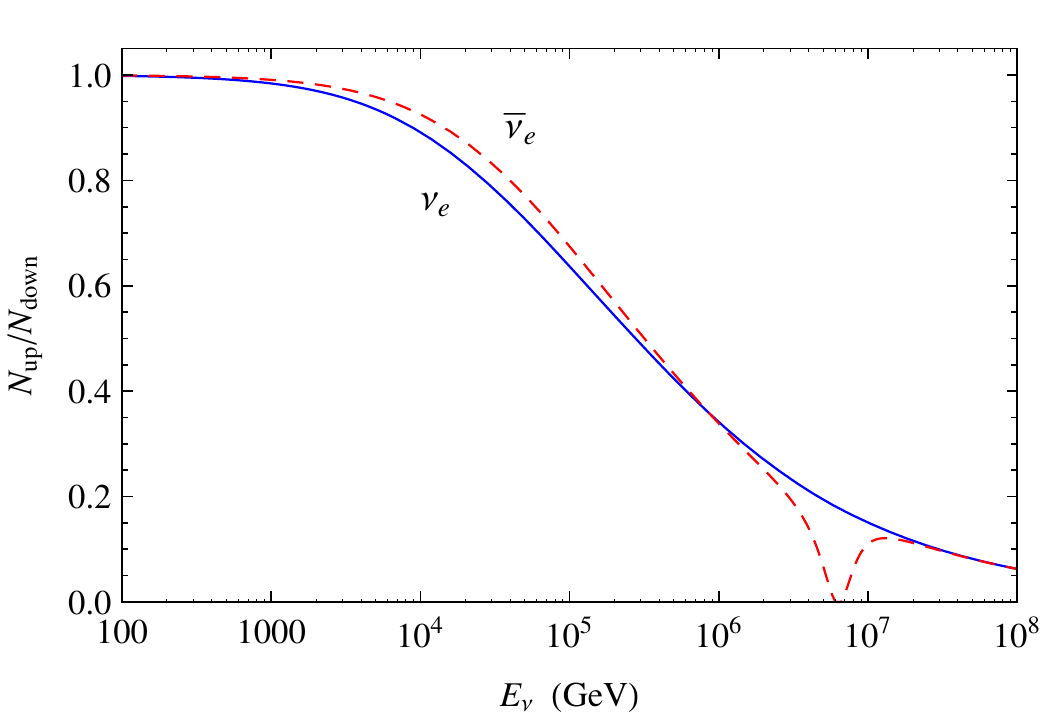}
\end{center}
\caption {\footnotesize
Ratio upgoing/downgoing events for 
$\nu_e$ and $\overline{\nu}_e$, plotted as a function of the neutrino energy. 
The calculation assumes isotropic fluxes, and integrates over the entire
up--going and down--going hemispheres.
\label{fig:abs_energy}
 }
\end{figure}

The angular distribution of the observable neutrino flux is
deformed because of absorption in the Earth, and one expects
a deficit for up--going neutrino events that have trajectories
that traverse the Earth.
Examples of the survival probability for 
$\nu_e$ and $\overline{\nu}_e$ as a function of zenith angle
for different neutrino energy are shown 
in fig.~\ref{fig:absorption}.  Integrating over the entire up--going and
down--going hemisphere with the assumption of isotropic fluxes,
one obtains the ratio   shown in fig.~\ref{fig:abs_energy}.

If one assumes  equal fluxes of $\nu_e$'s and $\overline{\nu}_e$'s 
and a spectral shape  of form $\propto E_\nu^{-2}$, after  integration 
in the energy range between 60~TeV--2~PeV, one estimates  a ratio
$N_{\rm down}/N_{\rm up} \simeq 1.84$. 
For softer spectra the ratio decreases, becoming 1.76 (1.70) for a 
power law spectrum of esponent 2.2 (2.4).

If the IceCube signal is  entirely formed by isotropic astrophysical neutrinos,
with a power law spectrum $\propto E^{-2}$, 
the probability 
to observe an up/down asymmetry as large or larger than the data 
is approximately 1.2\%. 
On the other hand one should take into
account the existence of  the estimated background of $6 \pm 3.4$ events
from atmospheric muons that is entirely down--going.
The observed up/down asymmetry is therefore additional evidence for an
important contribution of down--going events  generated by
atmospheric muons to the rate of contained events.

A possible explanation for the up/down asymmetry of the observations
is that the hypothesis of isotropy of the astrophysical flux is incorrect.
In fact if there is a significant contribution from sources
located in the Milky Way one would expect an excess of 
down--going events, because   from the South Pole  most of the Galaxy
(including the Galactic Center) is (always) above  the horizon.

It is  interesting to note that seven of the contained events have 
their origin in a $30^\circ \times 30^\circ$ region  
around the Galactic Center.  The probability of having one such a fluctuation
in the sky is of order 8\%;  nonetheless the  Galactic Center is a very
special location, and the result is certainly striking.
The assumption  that these events do indeed have their origin 
 from a region around the Galactic Center
implies that the neutrino emission  from this region has a luminosity 
(in the 60~TeV--2~PeV energy band) of order 
$L_\nu \simeq 5 \times 10^{36}$~erg/s. For a comparison 
the observations of the Galactic Center with the HESS 
gamma--ray telescope 
measure a photon luminosity above $E_\gamma = 1$~TeV:
$L_\gamma (E_\gamma > 1~{\rm TeV}) \simeq 7 \times 10^{34}$~erg/s, that is
approximately two order of magnitude smaller.

A Miky Way origin for a significant fraction
of the IceCube signal would therefore have very surprising
and exciting implications.

\section{Neutrino Induced muons}
\label{sec:muons}
The  background  due to down--going showers  
dominates the rate of contained events 
with an energy deposition below 60~TeV and, as discussed above,
has a non negligible rate above the threshold.
The IceCube data indicates that 
when the down--going background  is sufficiently small to allow
the  detection of  neutrino interactions as contained events,
the neutrino flux is already dominated by an astrophysical component
\footnote{The detection of shower events with energy in the 
range 80~GeV to 6~TeV has been performed using the DeepCore low
energy extension  of IceCube \cite{Aartsen:2012uu}.}

This situation is clearly not ideal, because there is not
the possibility to ``calibrate'' the observations using
the (relatively) well understood flux of atmospheric neutrinos
at lower energy.

Extending the measurement of the contained events to lower
energy one should  be able to observe the transition from
the standard atmospheric neutrino flux to the new component.
The transition should manifest itself as an hardening of the energy
spectrum  and as  a change in the angular distribution that evolves from the
the  characteristic ``secant law'' shape
of atmospheric neutrinos with $E_\nu \gtrsim 1$~TeV
to a shape determined by the distribution of the astrophysical sources
(an isotropic distribution if extragalactic emission is dominant).

The transition from a flux dominated  by
standard  atmospheric neutrinos to an astrophysical flux 
should also be observable using neutrino induced muons.
A $\nu_\mu$ ($\overline{\nu}_\mu$) of energy $E_\nu$, interacting
outside the detector generates
a flux of $\mu^\mp$ with an energy spectrum that extends 
to the maximum energy $E_{\mu , {\rm max}} \simeq E_\nu$. 
The study of  neutrino  indiced muon events  has already
allowed the measurement \cite{Abbasi:2010ie} of the atmospheric 
muon neutrino flux  for energies below 400~TeV using the
IceCube--40 data  (collected with the detector in construction),
and has also  allowed to set stringent limits
\cite{Abbasi:2011jx,Schukraft:2013ya}  on the size of a 
prompt or astrophysical contribution to the neutrino flux.

\begin{figure} [ht]
\begin{center}
\includegraphics[width=14.5cm]{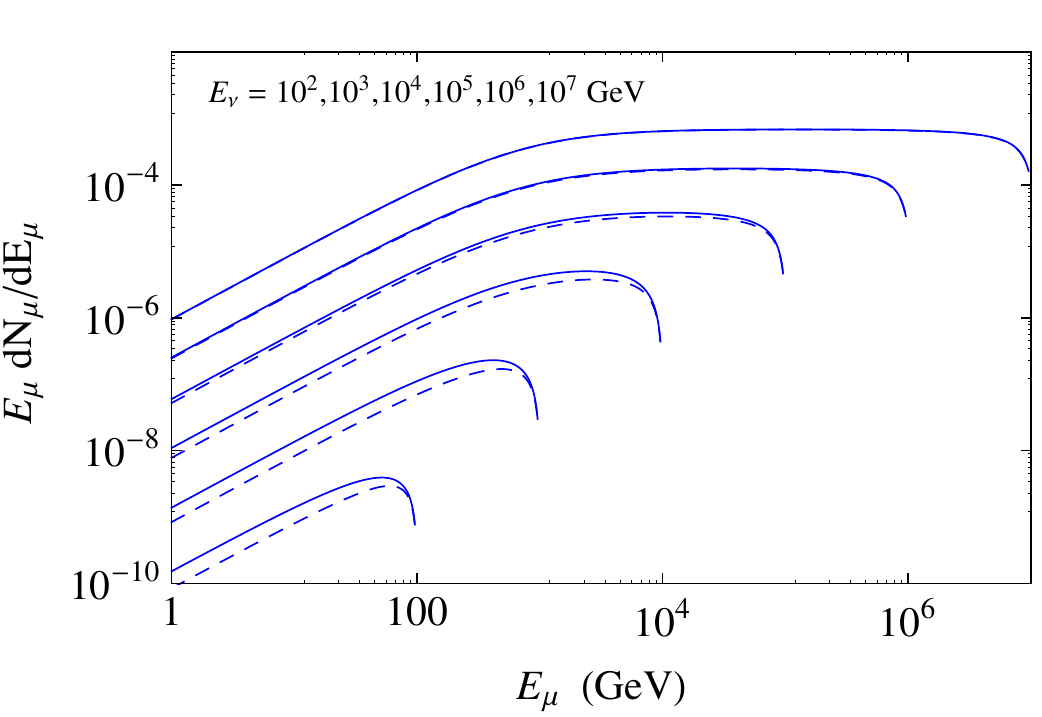}
\end{center}
\caption {\footnotesize
Differential muon yield for 
$\nu_\mu$ 
(solid lines) and 
$\overline{\nu}_\mu$ (dashed lines) of different energies.
\label{fig:muon_yield}
 }
\end{figure}

Some examples of the  differential muon yield
(that is the number of muons  generated per (anti)--neutrino) 
are shown in fig.~\ref{fig:muon_yield}.

\begin{figure} [ht]
\begin{center}
\includegraphics[width=14.5cm]{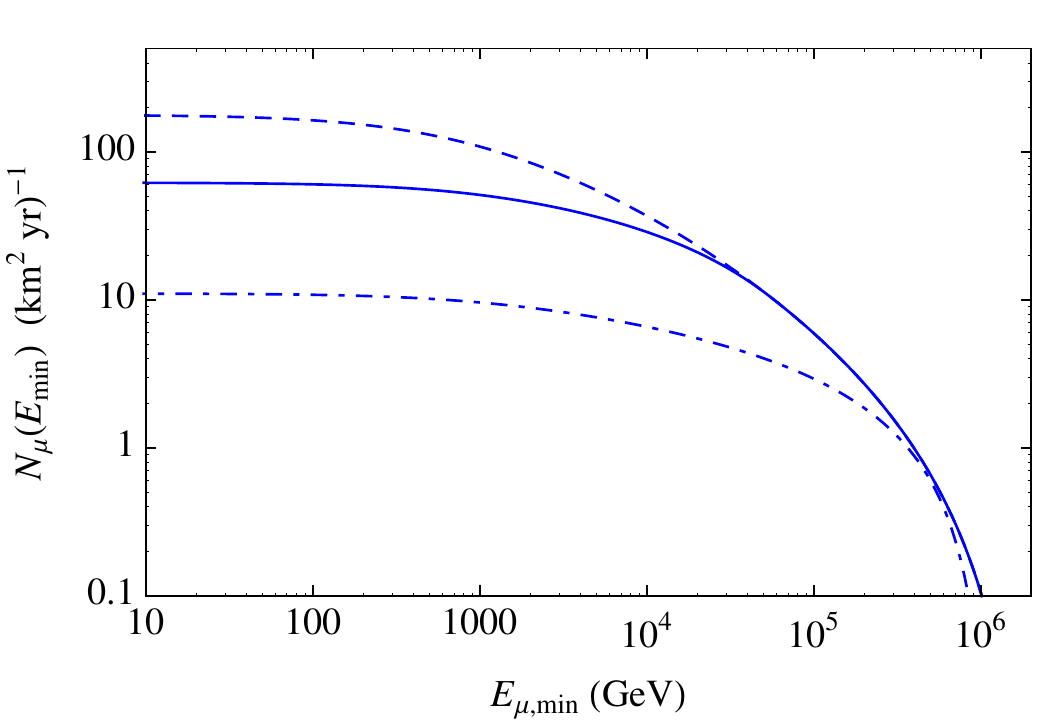}
\end{center}
\caption {\footnotesize
Neutrino induced muon flux integrated in solid angle in the 
 up--going hemisphere and in energy for $E_\mu \ge E_{\rm min}$.
The flux is calculated assuming equal isotropic fluxes
 of $\nu_\mu$ and $\overline{\nu}_\mu$.
The solid line corresponds to the neutrino flux
$E_\nu^2 [\phi_{\nu_\mu} (E_\nu) + \phi_{\overline{\nu}_\mu} (E_\nu)] 
 = 1.2 \times 10^{-8}$~ GeV/(cm$^2$\, s\, sr) between energies
60~TeV and 2~PeV.
The dashed line is calculated extending the same functional form
to a minimum energy 100~GeV. The dot--dashed line 
is calculated assuming that the neutrino 
spectrum is a narrow line at 1~PeV normalized 
to produce 1 contained event per year (with equal fluxes for all flavors).
\label{fig:muon_int}
 }
\end{figure}

Convoluting the muon yield with the neutrino fluxes (and
the survival probability in crossing the Earth) 
one can compute the observable flux of neutrino induced muons.
In figure~\ref{fig:muon_int} 
we show the integral flux of muons calculated 
using three models for the neutrino flux.
In the first one (motivated by the detection of the two events
with  PeV  energy   reported in \cite{Aartsen:2013bka} 
we have assumed only the existence of a narrow ``line''
of neutrinos  at $E_\nu \simeq 1$~PeV, normalized to produce two events
(assuming equal fluxes for all six neutrino types).
In the second model we used the flux estimated by IceCube
to explain the 28 events and given in equation (\ref{eq:flux-evidence}),
 assuming its validity in the range 60~TeV--2~PeV.
In the third model we extended the flux 
of equation (\ref{eq:flux-evidence}) to a minimum energy of 100~GeV.
The fluxes of neutrino induced muons 
integrated in angle over the entire  up--going hemisphere 
and in energy  for $E_\mu \ge E_{\rm min}$
are  
\{2.9, 5.9, 5.9 \}  events/(Km$^2$~year)  for a threshold $E_{\rm min} = 100$~TeV.
For the lower threshold  of  $E_{\rm min} = 10$~TeV,
the fluxes (in the same units)  become 
\{6.5, 29, 37 \}.
Lowering the threshold to 1~TeV and to  10~GeV   
the fluxes  are \{9.6, 51, 108 \}  and
 \{10 , 60, 164 \} respectively.
These predicted fluxes should be observable with the existing
IceCube data.  

\begin{figure} [ht]
\begin{center}
\includegraphics[width=14.5cm]{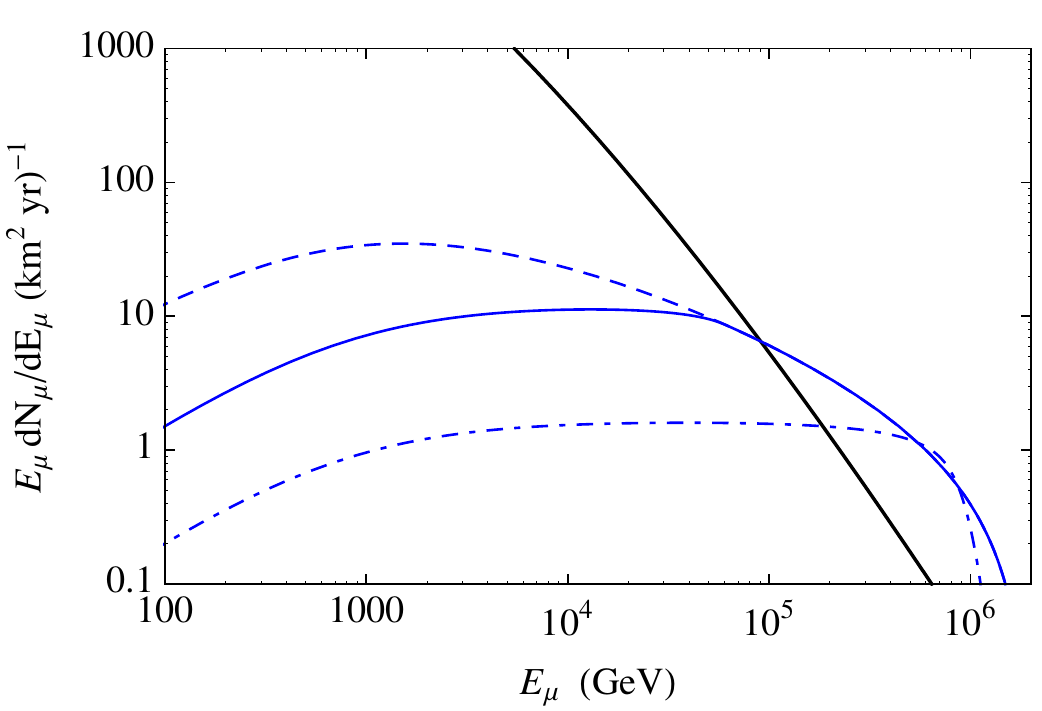}
\end{center}
\caption {\footnotesize
Neutrino induced muon flux as in fig.~\ref{fig:muon_int}
(with the same meaning  for the  different lines).
The flux is shown in differential form.
The thicker line describes atmospheric neutrinos.
 \label{fig:flux_muons}
 }
\end{figure}

The fluxes of neutrino induced muons  discussed above
are also shown in  differential form in fig.~\ref{fig:flux_muons} 
together with a prediction of the muon flux generated 
by standard atmospheric neutrinos. 
The extra--component should  become  visible as 
an hardening of the  muon spectrum  at 
high energy.  The hardening should also be associated
to a change in the angular distribution of the flux.
In fact the sensitivity to the existence of 
a  hard component in the flux could  be enhanced
selecting a smaller vertical angular region.
Such an angular  cut should  reduce more strongly
the flux of standard atmospheric neutrinos that has a maximum
for horizontal directions. 

The detection of  a new component of 
the neutrino flux  using  two  independent, different  methods
(contained  events and up--going muons)
would  strongly strengthen the  confidence in the result.

\section{Flavor composition}
\label{sec:flavor}
An often discussed method to identify the astrophysical origin
of a neutrino signal is a measurement of its flavor composition.
In general the flux of neutrinos of flavor $\alpha$
can be expressed in the form
\begin{equation}
\phi_{\nu_\alpha} (E,\Omega) = 
\sum_\beta
\phi_{\nu_\beta}^{(0)} (E,\Omega) 
~\left \langle P_{\nu_\beta \to \nu_\alpha} (E,\Omega) \right \rangle
\label{eq:oscnu}
\end{equation}
where the fluxes $\phi_{\nu_\beta}^{(0)} (E,\Omega)$ are the expected fluxes 
in the absence of neutrino oscillations
(determined by the flavor of the neutrinos at their creation point)
and $\left \langle P_{\nu_\beta \to \nu_\alpha} (E,\Omega) \right \rangle$ is the
oscillation probability averaged over the distribution of distance
of the neutrino production point.

In the energy range that we are considering here,
 flavor oscillations
are negligible for atmospheric neutrinos since the
shortest neutrino oscillation length 
(determined by the largest neutrino squared mass difference) is of order
$L_{\rm osc} \simeq (160~R_\oplus)~E_{\rm TeV}$, 
much longer than the Earth diameter. 
On the other hand  for astrophysical neutrinos the range 
of distances of the sources 
is likely to be much larger than the oscillation lengths,
so that averaging over distance in equation (\ref{eq:oscnu}) one
has:
\begin{equation}
\left \langle P_{\nu_\beta \to \nu_\alpha} (E,\Omega) \right \rangle =
\sum_j 
\left |U_{\alpha j}\right |^2
~\left |U_{\beta j}\right |^2
\end{equation}
with $U_{\alpha j}$ the neutrino mixing matrix. 
This implies that for  any flavor composition of the 
neutrino emission  at the sources, one will observe 
at the Earth  large  fluxes for all three flavors,
including a large flux of tau (anti)--neutrinos.

In what is by far the most plausible scenario, where the
astrophysical neutrinos are created in 
the chain decay of  charged pions and kaons, one expects that the fluxes
of the three different flavors are in good approximation equal.
The ratio neutrino/anti--neutrino is then determined 
by the relative importance of $\pi^+$ and $\pi^-$ (and $K^+$ and $K^-$) 
production at the source.

\section{The nucleon flux} 
\label{sec:nucleon-flux}
Two elements enter the calculation of the atmospheric neutrino fluxes,
the first is the description of the primary cosmic ray fluxes, 
the second is the modeling of the development of their showers 
in the atmosphere.
For the purpose of calculating the neutrino fluxes,
in good approximation, it is sufficient to have a good
knowledge of the cosmic ray ``all nucleon'' flux 
$\phi_N (E_0)$ that includes the contributions of 
all free (protons) and bound nucleons with energy per nucleon $E_0$.
The nucleon flux can be calculated summing the
contributions of all nuclear species present in the primary CR flux:
\begin{equation}
\phi_N (E_0) = \sum_A A^2 \; \phi_A (E_0 \; A)~.
\label{eq:flux-nucleons}
\end{equation}
The weight factor $A^2$ takes into account the number of
nucleons in a primary particle and the Jacobian factor 
between the energy per nucleon $E_0$ and the total energy of
the nucleus $E_{\rm tot} = A\, E_0$.
When the primary spectra are power laws
with approximately equal esponents 
($\phi_A(E) \simeq K_A \; E^{-\alpha}$),
the nucleon flux is also a power law with the same esponent,
and the contribution of nucleus $A$ is proportional 
to $K_A \; A^{-\alpha+2}$.
Since $\alpha > 2$, 
the contribution of bound nucleons is suppressed, and
free protons are the dominant component of the nucleon flux.

Because of the steepness of the cosmic ray energy
spectrum, neutrinos of energy $E_\nu$ are mostly produced
by primary nucleons with energy $E_0$ 
in a relatively narrow range above $E_\nu$.
Atmospheric neutrinos near 1~PeV are therefore mostly produced
in the showers of primary nucleons 
with energy just above the ``knee'',
a prominent softening feature in the CR spectrum 
at $E_{\rm tot} \simeq 3$~PeV, where 
the differential slope of the particle spectrum 
increases from 2.7 to approximately 3.

The knee in the CR spectrum has been known for many decades,
but its origin remains controversial and poorly understood.
Because of the smallnes of the flux,
cosmic rays in this energy range 
can only be measured in Extensive Air Shower (EAS) experiments, 
that are able to determine the mass number $A$ of the primary particle
with very poor resolution and large systematic errors.
EAS experiments can obtain a better measurement
of the total energy of a CR shower
(that depend only weakly from the mass number) 
and therefore usually only estimate an all particle flux 
(as a function of the total energy of the particle)
that sums over all nuclear types.
To infer the nucleon flux one has then the problem to model 
the composition of primary flux via
equation (\ref{eq:flux-nucleons}). 

\begin{figure} [ht]
\begin{center}
\includegraphics[width=14.5cm]{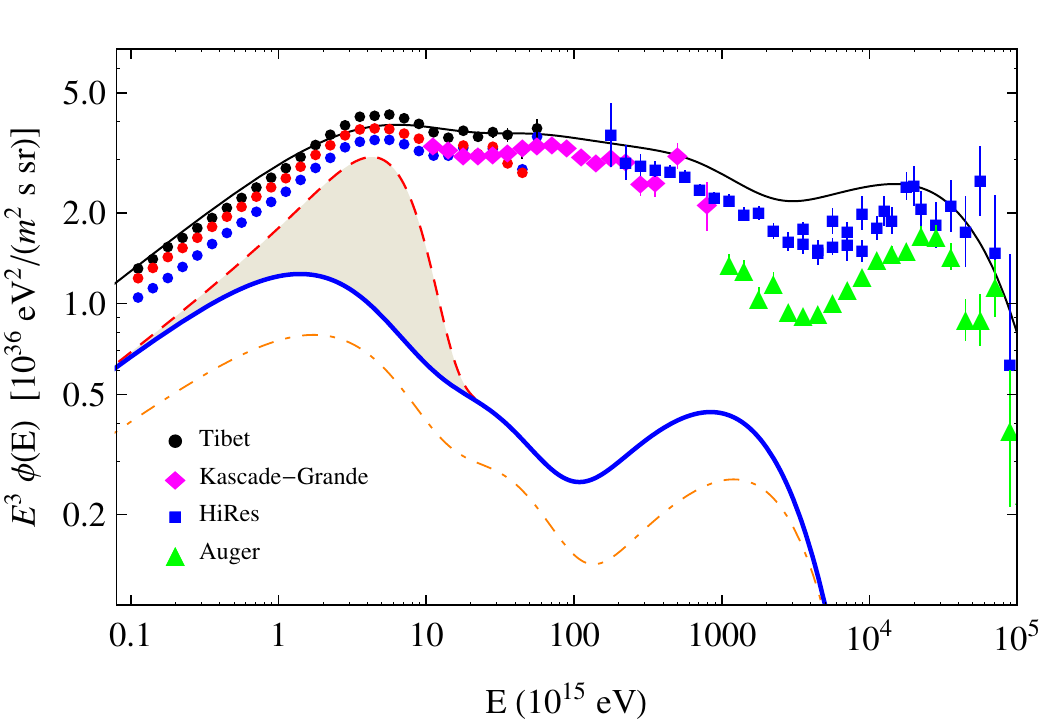}
\end{center}
\caption {\footnotesize
The points are measurements 
\cite{tibet,Apel:2012rm,AbuZayyad:2002sf,Abraham:2010mj}
of the CR all particle flux by the
TIBET III detector (circles, the three sets of data
are spectra estimated using different hadronic interaction models),
Kascade Grande (diamonds), HiRes (squares) and Auger 
(triangles).
The lines are the estimates of the all particle flux (thin, solid line)
the  proton flux (dot--dashed line) and the all nucleon flux 
(thick, solid line) of
Gaisser, Stanev and Tilav  \cite{Gaisser:2013bla}. 
The dashed line indicates a possible extra contribution 
to the all nucleon flux discussed in this work.
\label{fig:nucleon1}
 }
\end{figure}

\begin{figure} [ht]
\begin{center}
\includegraphics[width=14.5cm]{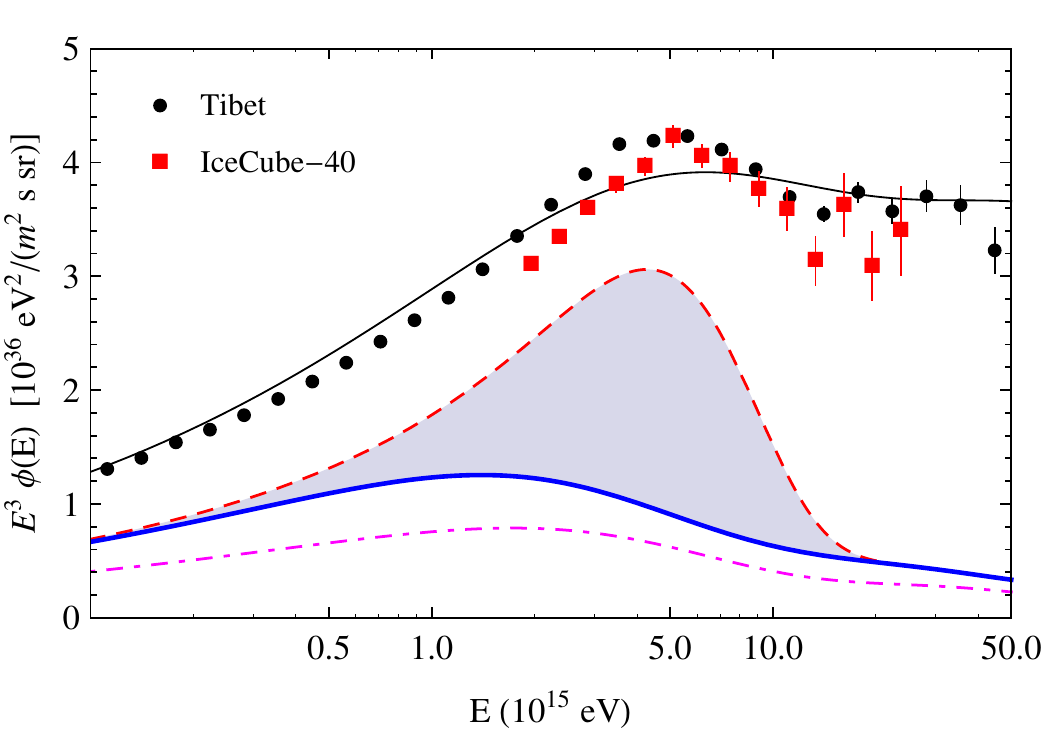}
\end{center}
\caption {\footnotesize
As in fig.~\ref{fig:nucleon1} (but with a linear scale for the
$y$ axis).  The squares are a measurement of the CR  flux  performed
by IceCube-40 \protect\cite{IceCube:2012vv}.
\label{fig:nucleon2}
 }
\end{figure}

Fig.~\ref{fig:nucleon1} and~\ref{fig:nucleon2} show a subset of the
available measurements of the CR all particle spectrum at high energy
obtained in EAS experiments.
The data are shown in the form of the product $\phi(E) E^3$ 
versus $E$ (with $E$ the particle energy) to enhance 
the spectral features. The $y$--axis scale is logarithmic 
in fig.~\ref{fig:nucleon1} and linear in fig.~\ref{fig:nucleon2}. 
The thin solid line is the  model  H3a of the CR spectrum by
Gaisser, Stanev and Tilav (GST) \cite{Gaisser:2013bla}. These authors 
discuss also a composition model of the CR spectrum 
that is decomposed into the sum of 5 groups of nuclei
(free protons, Helium, CNO, Mg--Si and Fe).
The (free) proton spectrum
of the model is shown in fig.~\ref{fig:nucleon1} 
and~\ref{fig:nucleon2} as the dot--dashed line, while the thick line
represents the all nucleon flux.
The dashed line in the figures is a ``toy model'' that 
we have constructed in the present work to describe a 
possible (albeit speculative) larger free proton flux.

This is not the place for a full critical discussion of the 
data and possible models of CR at the knee. The main point that we
would like to make here is that the data allow for an all nucleon
flux at the knee energy that is as much as three or four times larger than
the ``best fit'' as estimated in works such as GST without 
saturating the constraint that the nucleon flux must be smaller
than the particle flux.

This remark is not intended as a
criticism of the work of Gaisser, Stanev and Tilav,
who in fact carefully construct what, on the basis
of our present understanding, can be considered 
as a very reasonable ``best fit'' of high energy cosmic rays. 
The point is that we still have a poor understanding of 
what is the origin of the knee, and of the mechanisms that 
are responsible for its shape, and it is {\em possible} that 
the nucleon flux is significantly larger that ``baseline'' expectations.

The    estimate of the number of events in the
energy range 60~TeV to 2~PeV 
for  standard atmospheric neutrinos   discussed by IceCube 
 \cite{icecube-ICRC}    is  ($4.6 \pm 1.2$ events) and
has an uncertainty of order only 26\%.
Allowing more freedom for the composition of the CR primary flux
the uncertainty estimate   could very well be 
significantly larger.

The fundamental idea that uderlies the CR models of GST,
and that is in fact present in most of the works
that interpret the cosmic ray spectra,
is that the propagation and acceleration of cosmic rays
is controled by the particle rigidity 
(where the rigidity is the ratio $R = p/q$
with $p$ the momentum and $q = Z e$ the electric charge
of the particle), because particles with equal  rigidity have
identical trajectories in a magnetic field.
From this assumption it follows that features present 
in the proton spectrum should also be visible in the spectra 
of other nuclear species. 
A softening in the spectrum of (ultrarelativistic) protons 
at energy $E_p$ therefore implies a similar softening of the helium flux
at energy $2 \, E_p$, and more in general a softening 
at energy $Z \,E_p$ for the flux of nuclei with electric charge $Z e$. 
This idea, first discussed by Peters \cite{Peters} suggests that
the knee observed in the CR all particle spectrum 
should be understood as a sequence 
of softenings of the different nuclear species present in the CR flux.
This implies that the average mass of cosmic rays should increases
with energy.
Simultaneous measurements of different components 
of the CR showers at ground level,
in particular of the electromagnetic and muon components,
give information about the mass number of the primary particles.
This composition analysis depends on the modeling of 
hadronic interactions, but the results of Kascade \cite{Antoni:2005wq}
and other experiments have given clear indications
that indeed the CR mass composition becomes heavier above the knee
in a way that is consistent with the idea of a Peters sequence
of softenings. 
A detailed quantitative measurement of the evolution of
the composition is however not available.

It should be also noted that the idea that acceleration and propagation
are determined by the particle rigidity seems to imply that the slopes of 
the cosmic ray spectra of all particles types should be identical.
On the other hand there are clear indication that this is not the case,
in particular that the helium spectrum is harder than the proton spectrum,
so that the helium flux overtakes the proton spectrum well below the knee
(at $E_{\rm tot} \simeq 20$--50~TeV). The  difference in the slopes 
of the proton and Helium fluxes  has  crucial importance
for the extrapolation of the CR fluxes from the region of direct measurements
to the region of EAS.

A careful inspection of fig.~\ref{fig:nucleon1} and~\ref{fig:nucleon2}
shows that the shape of the energy spectrum around the knee 
is not well represented by a simple smooth softening. In fact, it appears 
that the knee is formed by a broad feature in the energy range 
 $E \simeq 2$--7~PeV, where the CR flux softens gradually, followed
by a sharper hardening a $E \simeq 15$~PeV, that has been clearly seen
by Kascade--Grande. These features are not well captured
by the hypothesis of a simple superposition of cutoffs in the
spectra of the individual components, and could be the indication of
 the need of a different type of description.

\section{Neutrinos from charm decay}
\label{sec:charm}
Most of the atmospheric neutrinos are produced in the weak decays
of charged pions and kaons (and at low energy also in the chain decay 
of the muons produced in association with the muon (anti)--neutrinos).
Charged pions and kaons have long lifetimes, and their decay becomes strongly
suppressed at high energy, when their decay length 
becomes longer that the interaction length.
Their decay probability decreases approximately $\propto E^{-1}$ because 
of relativistic effects.

Charmed particles have much shorter lifetimes, and 
decay rapidly with probability close to unity up to very high energy 
($E \sim 100$~PeV). 
For this reason, neutrinos from charm decay
become the dominant source of the atmospheric fluxes 
at a sufficiently large energy. 
A question that remains unanswered is if the transition energy
$E_{\rm charm}$ where the standard and charm decay contributions 
are equal is lower or higher than the energy $E_{\rm astro}$ 
where the flux of astrophysical neutrinos overtakes the atmospheric one.

The component of the atmospheric neutrino fluxes due to charm deacay
has properties in angular distribution, flavor composition and energy spectrum
that in principle allow its identification.

For $E_\nu \lesssim 30$~PeV the neutrino flux from charm decay is 
to a good approximation isotropic, reflecting the fact that nearly 
all charmed particles decay. In the case of neutrinos from pion and kaon decay
the angular distribution has a strong and characteristic dependence
on the zenith angle, because inclined showers develop higher
in the atmosphere, where the air density is lower and decay is more likely.

Charm decay generates
nearly identical fluxes of electron and muon (anti)--neutrinos, 
and a much smaller fraction of tau (anti)--neutrinos. This is the simple
consequence of the fact that in charm decay,
(with the exception of the 2--body leptonic decays of 
 $D^\pm_s$ mesons) there is no dynamical suppression for decays into
light charged leptons, and the mass difference between $\mu^\pm$ and $e^\pm$ 
results only in a small phase space suppression for muon neutrinos.
The $D^\pm_s$'s 
(bound states $[c\overline{s}]$ and $[s\overline{c}]$) have a
two body decay mode $\tau^+ \nu_\tau$ and $\tau^- \overline{\nu}_\tau$,
that, together with the chain decay of the $\tau^\pm$, are the source
of a smaller flux of tau (anti)--neutrinos.
The ratio $\overline{\nu}/\nu$ is determined by the relative importance
of the mechanisms $D\overline{D}$ and
$\Lambda_c \overline{D}$ in the charm production cross setion.
In the $D\overline{D}$ mode the $c$ and $\overline{c}$ charm 
(anti)--quarks are contained in two mesons, 
in the $\Lambda_c \overline{D}$ mode the $c$ quark is contained in 
a charmed baryon.
If the $D\overline{D}$ mode dominates, the ratio $\nu/\overline{\nu}$ is 
approximately unity, while
if the production of charmed baryon is not negligible
the ratio (that also depends on the shape of inclusive spectra of 
the charmed particles)   can be different from unity.
If the charmed baryons have an 
average energy significanly larger than the anti--charmed mesons,
it is possible to have a ratio $\nu/\overline{\nu}$ much larger than unity.

It is important to note that the flux of muon neutrinos
generated by charm decay  is accompanied by an
approximately equal (but slightly smaller)
flux of $\mu^\pm$. The muon flux is smaller because 
in the decay $c \to s \mu^+ \nu_\mu$ 
(or $\overline{c} \to \overline{s} \mu^- \overline{\nu}_\mu$)
the neutrino has a harder spectrum than the charged lepton 
because of the structure of the weak matrix element.
For the same reason in muon decay ($\mu^- \to \nu_\mu e^- \overline{\nu}_e$)
the electron has a harder energy spectrum than the $\overline{\nu}_e$.
This effect (that is neglected in several works, including Enberg et al.
\cite{Enberg:2008te}) amounts to a relative suppression of order 15--20\%
of the muon with respect to the muon--neutrino flux.
The muon flux can be   in principle measured to constrain
the size of the neutrino fluxes from charm  decay.

The energy spectrum of high energy neutrinos from pion and kaon decay 
with energy $E_\nu \gtrsim 10$~TeV is a steep power law
$\phi_\nu (E) \propto E^{-\alpha}$ with $\alpha \simeq \alpha_0 + 1$,
where $\alpha_0$ is the esponent of the primary nucleon flux.
This power law behaviour follows from the fact that the decay probability
of pions and kaons is to a good approximation $\propto E^{-1}$ and that
the inclusive spectra of particles in the forward region 
is approximately scaling (see for example Gaisser \cite{Gaisser-book}
or \cite{Lipari:1993hd}). 
In the case of neutrinos from charm decay the decay probability
suppression is absent, and the spectrum is harder.
The detailed shape of the spectrum,
as well as its normalization, are however model dependent. 
The important elements that enter the calculation
(together with the nucleon flux) are 
the energy dependence of the charm production cross section
$\sigma_{c\overline{c}}(s)$ 
and the inclusive spectra (and composition) of the charmed 
particles in the final state of hadronic interactions.

Discussing the results on contained events, the IceCube collaboration
\cite{icecube-ICRC} has argued that it is not possible to
explain the data as the effect of neutrinos from charm decay, because
a sufficiently large prompt neutrino flux 
is excluded by the data of IceCube--59 \cite{Schukraft:2013ya}.
This argument is constructed on the assumption
that the shape of the energy spectrum of the prompt component of
the neutrino flux is well determined, and that it is possible to 
change the normalization of this flux, leaving its spectral shape constant. 
We will argue in the following that this assumption is too restrictive, 
and that it is possible that the prompt neutrino flux is significantly harder
than what has been estimated by IceCube.

In this work we do not attempt a critical discussion of the theoretical
uncertainties in the calculation of the fluxes of neutrinos and muons
from the decay of charmed particles, and we do not review
the different calculations that are available in the literature
(for this purpose see for example Gaisser \cite{Gaisser:2013ira}).
We  also do not attempt to construct a ``best fit'' prompt neutrino
flux on the basis of a detailed model of the charm production cross section.
A recent example of such a calculation, performed in a perturbative QCD
framework, is the work of Enberg et al. 
\cite{Enberg:2008te} that has been used as a reference by the IceCube
collaboration.
Rather we argue  that it cannot be excluded that the charm production cross
section receives important contributions from (poorly understood)
non--perturbative mechanisms, resulting in a spectrum with a different
shape and normalization.
Using this purely phenomenological approach it could be possible
to explain the rate of contained event measured
by IceCube as the consequence of a large 
flux of prompt atmospheric neutrinos without violating
known physical principles and without entering in conflict with 
existing data.
Such a flux of prompt neutrinos is larger and harder than
the best motivated calculations, but it remains as a logical possibility.
It is clearly desirable to falsify this possibility {\em experimentally}.

To illustrate this point one can consider a very simple toy model
that takes into account charm production only in the first interaction
of a primary nucleon. The model is defined by the
cross section $\sigma_{c \overline{c}}(s)$, and by the 
inclusive energy spectra of the final state charmed particles.

Recently, measurements of the charm production cross section 
in $pp$ interactions at high energy 
have been obtained  at $\sqrt{s} = 200$~GeV by the PHENIX \cite{Adare:2010de} 
and STAR \cite{Adamczyk:2012af} collaborations
and at $\sqrt{s} = 2.76$ and 7~TeV  
by the ALICE collaboration \cite{ALICE:2011aa,Abelev:2012vra} 
at LHC. Lower energy measurements are reviewed in \cite{Lourenco:2006vw}.
At $\sqrt{s} = 200$~GeV (that corresponds to a lab. frame energy
$E_0 \simeq 20$~TeV) the PHENIX collaboration 
estimates the charm production cross section as 
\begin{equation}
\sigma_{c\overline{c}} = 551 \pm 57 \; ({\rm stat.}) \pm 193 \; ({\rm sys.}) ~\mu{\rm b}
\end{equation}
while the STAR collaboration measures:
\begin{equation}
\sigma_{c\overline{c}} = 797 \pm 210 \; ({\rm stat.}) ~^{+208}_{-295} \; ({\rm sys.}) ~\mu{\rm b}
\end{equation}
At LHC energies ALICE measures:
\begin{equation}
\sigma_{c\overline{c}} (2.76~{\rm TeV}) = 4.5 \pm 0.8 \; ({\rm stat})
~^{+1.0}_{-1.3} \; ({\rm sys})
~^{+2.6}_{-0.4} \; ({\rm extrapolation}) ~~{\rm mb}
\end{equation}
and
\begin{equation}
\sigma_{c\overline{c}} (7~{\rm TeV}) = 8.5 \pm 0.5\; ({\rm stat})
~^{+1.0}_{-2.4} \; ({\rm sys.})
~^{+5.0}_{-0.4} \; ({\rm extrapolation})
~~{\rm mb}
\end{equation}
This corresponds to a power law growth of the charm production cross
section $\sigma_{c\overline{c}} (s) \sim s^p$ with $p \simeq 0.6$--0.8.
It should however be stressed that the experiments measure the cross section
in a limited  region of phase space
(for ALICE the rapidity region $|y| < 0.5$), 
 that accounts for only a small fraction of the cross section
(of order 0.2 at 200~GeV and 0.1 at LHC) and then 
extrapolate to the entire phase space on the basis of theoretical models.
From a purely phenomenological point of view there is therefore 
considerable freedom in the description of the charm cross section in the 
kinematical forward region.

\begin{figure} [ht]
\begin{center}
\includegraphics[width=14.5cm]{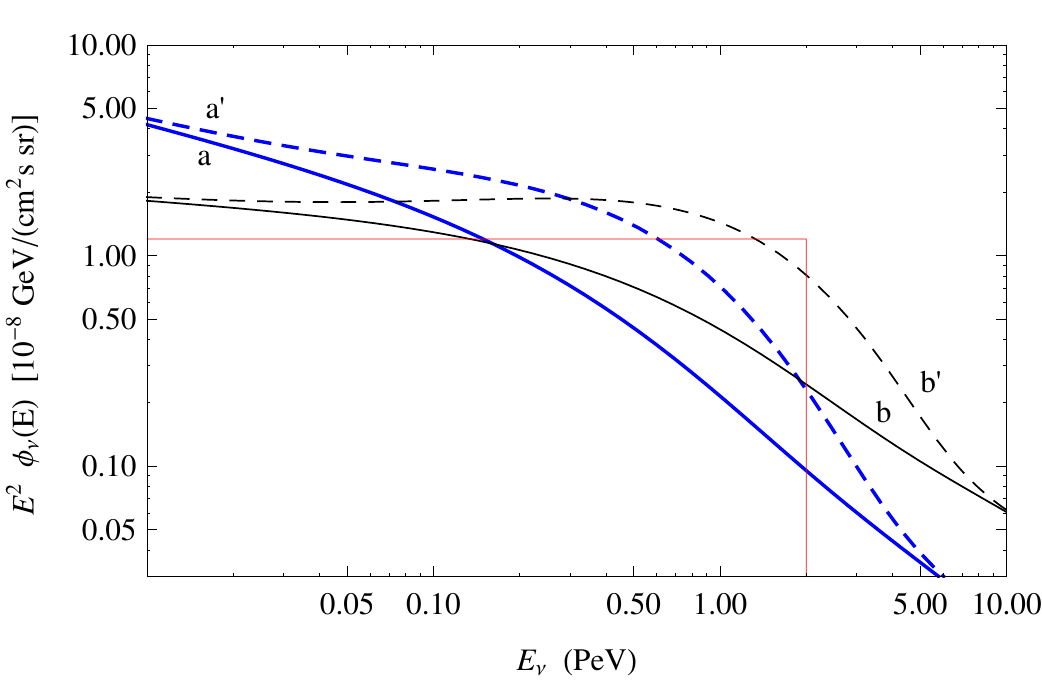}
\end{center}
\caption {\footnotesize The  figure shows the result of 
calculations of the atmospheric prompt neutrino  flux
($\nu_\mu + \overline{\nu}_\mu$ or
$\nu_e + \overline{\nu}_e$) 
performed with four very simple toy models.
For the solid lines labeled $a$ and $b$ the nucleon flux is the model H3a
of GST \protect\cite{Gaisser:2013bla}. 
The dashed lines are computed adding to the GST flux the extra component
to the all nucleon flux shown 
in fig.~\ref{fig:nucleon1} and~\ref{fig:nucleon2}.
See the main text for a description of the models.
The thin rectangular line gives the neutrino flux
indicated by IceCube as a possible source of the contained events.
\label{fig:charm}
 }
\end{figure}

To explore the possibility that a flux as large as 
the one indicated by the IceCube collaboration 
in equation (\ref{eq:flux-evidence})
could due to neutrinos from charm decay,
we have performed some calculations with very simple ``toy'' models.
Four examples of  such  calculations 
are shown in fig.~\ref{fig:charm}. Two calculations
(shown as the solid lines labeled $a$ and $b$) were performed
using the nucleon flux of the  model H3a
of GST \protect\cite{Gaisser:2013bla}, while in the other two
(shown as the dashed lines labeled $a^\prime$ and
$b^\prime$) we have added to the GST flux the extra component
of free protons shown in fig.~\ref{fig:nucleon1} and~\ref{fig:nucleon2}.
For the calculation $a$ and $a^\prime$ the charm cross section 
in $pp$ interactions was  assumed to be 
$\sigma_{c\overline{c}}(E_0) \simeq 0.7~[\sqrt{s}/(200~{\rm GeV})]^{0.7}$~mb,
with a scaling  inclusive cross section for the charmed particles
$\propto (1- |x_F|)^3$  (with  $x_F$ the Feynman $x$  variable).
For the calculations of lines labeled
$b$ and $b^\prime$ the charm cross section
was $\sigma_{c\overline{c}}(E_0) \simeq 0.08~[\sqrt{s}/(200~{\rm GeV})]^{1.2}$~mb,
and a scaling inclusive cross section $\propto (1- |x_F|)$.
Note that this cross section is smaller than the detected one,
and therefore must be understood as only a 
fraction of the charm cross section
that corresponds to the production of charmed particles in the fragmentation
region. All these particles were treated as $\Lambda_c$'s.
For simplicity we have only taken into account charmed particles produced
in the first interaction.

The toy models that we have just discussed are of course
very naive and speculative, but indicate  that 
it seems possible to obtain  a prompt neutrino  flux 
as large as the one needed to explain the 
IceCube contained events  with neutrinos from charm decay.
This requires assumptions on the dynamics of charm production
that  do not correspond to the presently favored
theoretical models,  but are not manifestly impossible.

\section{Outlook}
\label{sec:outlook}
The announcement of the detection of two contained events with
PeV energy by IceCube, and more recently of a larger set 
of contained events with visible energy in the range between 
60~TeV and 1~PeV has  understandably been met with great interest. 
Several authors have immediately started the discussion 
on the implications of these results  when interpreted as
the manifestation of a flux  of astrophysical neutrinos.

An  astrophysical origin for a large  fraction of the 
contained  events is probably the simplest
and most natural explanation of  the data, but 
(conservatively) an instrumental
or atmospheric origin cannot  be entirely ruled out.

From a purely experimental point of view the most important 
limitation of the IceCube result is that,
because of the background of down--going showers,
the detection of contained neutrino events  has been possible
only at very high energy ($E_{\rm vis} \ge 60$~TeV) where the 
event rate seems to be already dominated by an astrophysical flux.
This does not allow the direct observation of 
the required transition from a neutrino flux
dominated by atmospheric neutrinos to the harder flux generated by
astrophysical sources. 

The direct observation  of this  transition 
is clearly very desirable and  should be possible studying the
flux of neutrino induced muons that enter the detector
with up--going trajectories \cite{Abbasi:2011jx,Schukraft:2013ya}.

The lack of a positive detection of a harder flux component in these studies
is not in contradiction with the contained event analysis,
but if the contained events are produced by real neutrino interactions
a new harder component  in the neutrino flux should  soon be 
detected with  in up--going  muon events.

In this  work we have also investigated the possibility  that
the contained  events observed by IceCube are  generated
by  atmospheric neutrinos. We have demonstrated   that this 
requires assumptions  that are  not the simplest and most natural,
but that are not impossible or in contradiction with existing data.

For atmospheric neutrinos with  PeV energy  
(the crucial  energy range   investigated here), 
a large uncertainty is due to the fact that they are generated by
primary particles at or just  above the ``knee'' in the CR spectrum,
a structure  that remains  poorly understood.
Modifying the composition of the CR  near the knee there   is room
to increase  (or in fact also decrease)  the 
neutrino  flux by  a large factor.

Very recently, our understanding  of 
the charm production cross section has made great 
progress, but  we are still lacking  in data  on the  size and
properties of the cross section in the fragmentation region. 
If  charm production in the  forward region of the interaction
kinematical   space  is larger  than  expected and
grows with  c.m. energy in an appropriate way, it could be possible
to obtain the observed rate of  contained events 
with  neutrinos  from charm decay.

The bottom line of this  discussion is that it is  clearly very desirable
to exclude the speculative possibilities  considered  above
 not  with theoretical arguments  but experimentally.
The  detection of  a  large  fraction  of tau neutrinos
in the  flux  would be unambiguous  evidence for  the 
presence of astrophysical neutrinos.
This detection could  be accomplished with the identification
of  tau  decays  inside the detector,  but also with a careful
comparison of the rates of   shower and track  contained events,
or of the rate of contained events   with   the fluxes of 
neutrino induced muons.
Also a   comparison of the neutrino and (down--going) muon fluxes 
could give information on the size of the 
prompt neutrino  component.
This  method however suffers  from   the  problem of a
possible   contamination of muons
generated in the decay of flavorless mesons  \cite{Illana:2010gh}.

\vspace{0.3 cm}
\noindent {\bf Acknowledgments}   
I would like to   thank
Tom Gaisser, Spencer Klein and
Todor Stanev for   stimulating discussions and clarifications about
the IceCube results.

\clearpage

\end{document}